\documentclass[pdflatex,sn-mathphys-ay]{sn-jnl}

\usepackage{graphicx}%
\usepackage{multirow}%
\usepackage{amsmath,amssymb,amsfonts}%
\usepackage{amsthm}%
\usepackage{mathrsfs}%
\usepackage[title]{appendix}%
\usepackage{xcolor}%
\usepackage{textcomp}%
\usepackage{manyfoot}%
\usepackage{booktabs}%
\usepackage{algorithm}%
\usepackage{algorithmicx}%
\usepackage{algpseudocode}%
\usepackage{listings}%
\usepackage{aas_macros}%

\def\tvi(#1,#2){\vrule height #1pt depth #2pt width 0pt}
\def\p{\partial}
\def\e{{\rm e}}
\def\d{{\rm d}}

\def\rso{r_{\rm s}}

\def\M{{\cal M}}

\theoremstyle{thmstyleone}

\theoremstyle{thmstyletwo}

\theoremstyle{thmstylethree}

\begin{document}

\title[Impact of rotation of the accretion of entropy perturbations]{Impact of rotation on the accretion of entropy perturbations in collapsing massive stars}

\author*[1]{\fnm{Olzhas} \sur{Mukazhanov}}\email{21210190055@m.fudan.edu.cn}

\affil*[1]{\orgdiv{Department of Physics}, 
           \orgname{Fudan University}, 
           \orgaddress{\city{Shanghai}, \postcode{200438}, \country{China}}}
           
\abstract{Convection in the innermost shells of massive stars plays an important role in initiating core-collapse supernovae. When these convective motions reach the supernova shock, they create extra turbulence, which helps energize the explosion. In our earlier work, we studied the effect of rotation on the hydrodynamic evolution of convective vortices in collapsing stars. This study focuses on how rotation influences the entropy perturbations, which naturally form in turbulent convection. As these perturbations are carried inward with the collapsing star, they generate both vorticity and sound waves. Using linear perturbation theory, we model entropy waves as small disturbances on top of a steady background flow. Our results show that stellar rotation has little effect on the evolution of entropy perturbations during collapse, prior to encountering the supernova shock. This outcome is consistent with our earlier findings on the limited influence of rotation in the accretion of convective eddies.}

\keywords{accretion, accretion discs, convection, hydrodynamics, massive stars, supernovae, turbulence}

\maketitle

\section{Introduction}

Core-collapse supernovae mark the violent deaths of massive stars, releasing vast amounts of energy and driving the synthesis of heavy elements \citep[e.g.,][]{burrows:13a, janka:16a}. When the stellar core exhausts its nuclear fuel, it becomes unstable and collapses under gravity, forming a neutron star or black hole \cite[e.g.,][]{oconnor11, kotake:12snreview, Burrows25}. The collapse launches a shock wave that initially stalls, and its subsequent revival through neutrino heating, turbulence, and rotation remains a central open problem in astrophysics \cite[e.g.,][]{foglizzo:00, mueller20b, Mezzacappa20}. Understanding these explosions is essential for linking stellar evolution with compact object formation, nucleosynthesis, and multimessenger signals such as neutrinos and gravitational waves \citep[e.g.,][]{Nakamura16}. 

One of the effects that may help the explosion is the role of convection in pre-collapse stars. The progenitors of core-collapse supernovae develop strong convection in their nuclear-burning layers prior to iron core collapse \citep[e.g.,][]{arnett:11b, Fields21, Varma21}. This activity is thought to play an important role in shaping the supernova explosions \cite[e.g.,][]{couch:13d, couch:15b,mueller:17}. After the iron core collapses, the convective shells fall toward the center \citep[e.g.,][]{lai:00, takahashi:14, Boccioli25}. The supernova shock, formed at core bounce, collides with these perturbations \citep[e.g.,][]{mueller:15,mueller:16b}. This interaction enhances the chaotic, non-radial motions in the region behind the shock \citep{mueller:17, huete:19}. As a result, extra pressure is produced that helps the shock expand and makes the conditions for an explosion more favorable \citep{takahashi:16, nagakura:19, Vartanyan22, Bollig21}. Among the different shells, the innermost shells, those undergoing oxygen and (to a lesser extent) silicon burning are, expected to contribute most strongly to shaping the explosion conditions \citep{collins:18}. See, e.g., \citet{mueller:20review} for a recent review. 

Using linear hydrodynamic equations, \citet{abdikamalov20} investigated how convective vortices evolve during the collapse of non-rotating stars under idealized conditions for non-rotating stars. A simplified framework enables systematic parameter studies and helps to elucidate the key physical mechanisms \citep[e.g.,][]{fernandez:09a, foglizzo:09}. They showed that accelerating collapse stretches vortices radially, reducing their velocity, while at the same time distorting density surfaces and inducing pressure fluctuations that generate acoustic waves \citep{kovalenko:98, mueller:15}. When these disturbances reach the shock, velocity perturbations from acoustic waves are expected to dominate over those from the vortices themselves \citep{huete:18}. 

Some stars are expected to be rotating \citep[e.g.,][]{Heger:05, popov:12, mosser:12, deheuvels:14, Dhanpal25}. In \citet{abdikamalov21} (hereafter Paper I), we investigated the role of rotation in the fluid-dynamical evolution of convective vortices over the course of stellar collapse. Employing first-order linear perturbation theory within a simplified framework, we traced the vortices from convective shells down to the vicinity of the supernova shock. Our main conclusion was that rotation has only a minor effect on their evolution. The reason is that, prior to shock interaction, i.e. when the perturbations are still at large radii, the infalling matter does not spin up sufficiently to alter their dynamics. This finding holds even for the most rapidly rotating progenitors. Building on this, the present work focuses instead on the influence of rotation on entropy perturbations, which are generated by turbulent convection in nuclear-burning shells of massive stars \citep[e.g.,][]{yadav20, Yoshida21}.

Entropy perturbations in stellar matter generate vorticity during collapse through the baroclinic effect \citep{foglizzo:01}. In a non-uniform flow with entropy variations, surfaces of constant pressure do not coincide with those of constant density. As a result, the net pressure force on a fluid element does not pass through its center of mass, creating a torque that produces additional vorticity, as explained in more detail below. Both entropy and vorticity waves can also excite acoustic waves as they propagate: vorticity waves, in particular, distort density surfaces and thereby induce pressure perturbations \citep[e.g.,][]{foglizzo:06, foglizzo:07}.
In this work, we examine how rotation influences these processes. As shown below, the effect of rotation is found to be weak. The reason is straightforward: before encountering the shock, when the matter is still at large radii, the spin-up during infall is insufficient to alter the dynamics in a significant way.

A clarification is in order. Although stellar matter does not spin up to dynamically important rotation rates before shock crossing, rotation can still play an important role in the post-shock region and give rise to a variety of interesting effects in rapidly rotating stars \citep{takiwaki:16, Summa18Rotation, Piran19jet_ccsn, Shishkin21Supplying}. For instance, centrifugal deformation at core bounce can produce significant gravitational-wave emission \citep{Zwerger97, Fuller15, mitra23, mitra24, Abylkairov24} and lead to non-axisymmetric modes developing after core bounce \citep[e.g.,][]{Shibagaki20new}\footnote{Gravitational waves are also emitted in slowly rotating or non-rotating models, where hydrodynamic instabilities contribute to the signal (see \citet{Mezzacappa24review} for a recent review).}. For extremely rapid rotation, magnetic fields become increasingly important \citep{burrows:07b, winteler12, Raynaud20, Bugli23}. They can transport angular momentum from the proto-neutron star to the shock front, depositing energy and driving powerful hypernova explosions \citep{bisno:76, akiyama:03, moesta:14a, kuroda:20, Shibagaki24}. In some scenarios, such explosions may also give rise to long gamma-ray bursts (GRBs) \citep{woosley:06, metzger11protomagnetar, obergaulinger:20}. GRBs are highly collimated, jet-driven explosions \citep{Sari99jets}, most often detected when viewed on-axis \citep[e.g.,][]{Beniamini19}, though off-axis detections are also possible \citep{Granot02, Nakar17, Abdikamalov25}. They have been observed across the electromagnetic spectrum \citep{Kumar15Physics}, including in the optical band \citep{Vestrand05Link, Klotz09, Komesh23, Abdullayev25}. 

The paper is organized as follows. Section \ref{sec:method} describes the method, Section \ref{sec:results} presents the results, and Section \ref{sec:conclusion} summarizes the conclusions.

\section{Methdods}
\label{sec:method}

As in Paper I, we model entropy perturbations as linear perturbations on top of the mean flow of the collapsing star. The stellar material is described by an ideal gas equation of state with adiabatic index $\gamma=4/3$, a reasonable representation for for stellar matter dominated by radiation \cite[e.g.,][]{fernandez:15a}. A rotating transonic Bondi solution is used to model the background flow. It exhibits a sonic radius $\rso$, where the flow transitions from the subsonic to the supersonic regime. Assuming constant specific angular momentum radial profile, the solution depends on $\gamma$, the core mass, rotation, and pressure. We take the central mass as $1.4M_\odot$, which is in line with the mass of proto neutron stars obtained in modern 3D simulations \citep[e.g.,][]{nagakura20, mueller25minimum}. Since $\rso$ depends on the sound speed, and hence on temperature, setting its value fixes the pressure scale. We adopt $\rso=1.5 \times 10^3\,\mathrm{km}$, a typical value for the early phase of post-bounce evolution \citep[e.g.,][]{takahashi:14}. This model accounts for the leading-order effects of rotation: the differential rotation caused by angular momentum conservation and the slowdown of collapse due to centrifugal force. Details of the method used to obtain the background solution can be found in Appendix B of Paper I.

We explore ten different values of specific angular momentum $L$, spanning from 0 up to $L_\mathrm{max} = 3\times10^{16}\,\mathrm{cm^2/s}$. This upper value exceeds by a factor of a few the maximum angular momenta expected even in the most rapidly rotating pre-supernova cores, which are estimated to be ${\lesssim}10^{16}\,\mathrm{cm^2/s}$ \citep[e.g.,][]{woosley:06}. We nevertheless adopt it as an extreme case to probe the effect of rotation over a broad range. As shown below, even at this extreme limit, rotation has little influence on the perturbations during collapse, prior to their interaction with the supernova shock.

Following our Paper I, the perturbation dynamics are described by linear hydrodynamic equations, which we express as a 2nd-order ordinary differential equation (cf. Appendix~\ref{sec:formalism} for its derivation),
\begin{flalign}
\left\lbrace{\p^2 \over \p X^2}+W\right\rbrace( r\delta \tilde \upsilon_\phi)=-
\e^{\int {i\omega'\over c^2}\d X}{\p \over \p X}{r\delta w_\theta\over \upsilon_r}, \label{rdvtilde1} 
\end{flalign}
where $X$ is related to $r$ coordinate (via Eq.~\ref{eq:X}), while $\delta \tilde \upsilon_\phi$ corresponds to the $\phi$-velocity perturbation (via Eq.~\ref{eq:vphitilde}). We carry out the solution in spherical coordinates, restricted to the equatorial plane, as outlined in Appendix~\ref{sec:formalism}. To simplify, we neglect poloidal derivatives \citep{Walk23}. The angular dependence is separated by assuming a $\exp(im\phi)$ form, where $m$ stands for the azimuthal wavenumber, and dependence on time is separated as $\exp(-i\omega t)$, with $\omega$ representing the frequency. In this formulation, $\delta w_\theta$ is the vorticity component in the $\theta$-direction, which represents vortex motion confined to the equatorial plane. We require no incoming acoustic waves from infinity \citep{abdikamalov20}. We obtain a homogeneous solution using the Frobenius expansion and imposing regularity at the sonic radius \citep{foglizzo:01}. See Appendix~\ref{sec:formalism} for the details of the method. This method has been validated through comparison with nonlinear hydrodynamic simulations by \citet{telman24}, who demonstrated that the perturbative analysis successfully reproduces all key features of convective vortex evolution in collapsing stars.

We note that our method, though simplified, is deliberately designed to isolate the influence of rotation from other physical processes. While not intended to yield precise quantitative predictions, it effectively highlights the underlying mechanisms and enables systematic parameter studies, thereby complementing large-scale simulations. 

\section{Results}
\label{sec:results}

\begin{figure}
    \centering
    \includegraphics[width=0.49\textwidth]{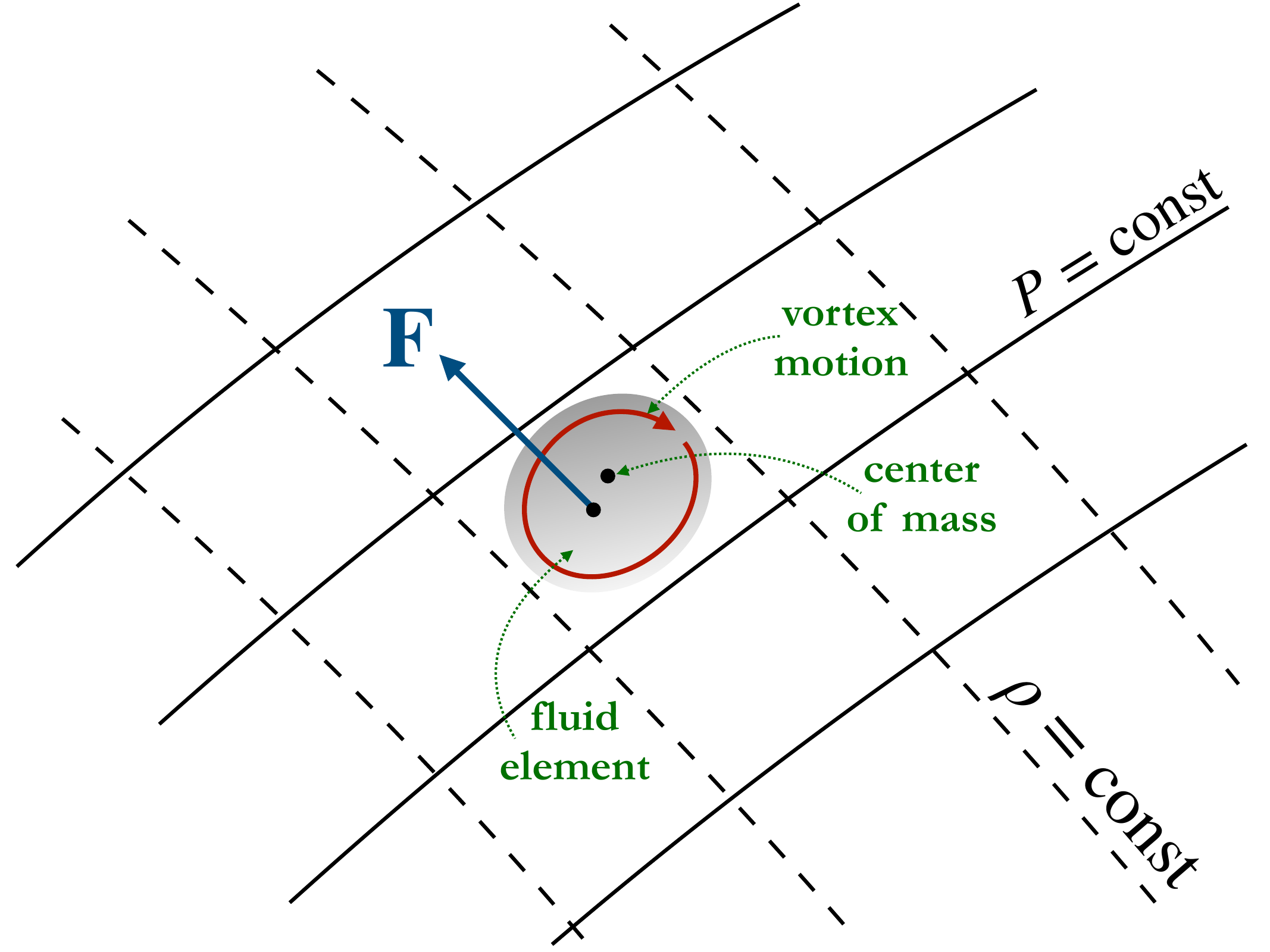}
    \caption{Schematic depiction of generation of vorticity by entropy perturbations in non-uniform flow. When the surfaces of constant pressure (solid lines) do not match with the surfaces of density (dashed lines), the total pressure force on a fluid element does not pass through is center of mass. This exerts a torque, generating a vortex motion.}
    \label{fig:ent_scheme}
\end{figure}

Section \ref{sec:qualitative} below provides a qualitative overview of the physical process, while Section \ref{sec:quantitative} explores it in greater quantitative detail.

\subsection{Qualitative Picture}
\label{sec:qualitative}

If the entropy variation in a star exhibit a non-radial structure, the surfaces of constant density no longer coincide with those of constant pressure. Consequently, the pressure force acting on a fluid element is not directed through its center of mass. This misalignment gives rise to a baroclinic effect, which exerts a torque and thereby generates vorticity \citep[e.g.,][]{foglizzo:01}. The mechanism is schematically illustrated in Fig.~\ref{fig:ent_scheme} \citep[see, e.g.,][for a pedagogical explanation]{ThorneBlandford17}. As this vorticity is advected inward, it perturbs the flow and distorts the iso-density surfaces, which in turn leads to the emission of acoustic waves \citep{mueller:15, abdikamalov20}. 

In addition, entropy perturbations can also couple directly to the background flow to produce acoustic waves. Consider a fluid element with mass $m$ with entropy variation of $\delta S$. How much the element expands under a change in pressure depends on its entropy. This expansion produces sound waves. If the element moves from a region with average specific enthalpy $h_1$ to another region with enthalpy $h_2$, the energy of the sound waves is \citep{foglizzo:00}
\begin{equation}
\delta E \sim (h_2-h_1) \delta m
\end{equation}
where $\delta m$ is the change in the mass $m$ of a fluid element that keeps the same volume but has an entropy perturbation of $\delta S$,
\begin{equation}
\label{eq:deltam}
\delta m = m \frac{\delta \rho}{\rho} = m \frac{\delta S}{\gamma c_\mathrm{v}},
\end{equation}
where $c_\mathrm{v}$ is the specific heat at constant volume. The specific energy carried by the emitted sound waves is therefore \citep{foglizzo:00}:
\begin{equation}
\label{eq:e21sound2}
\delta {\cal E} \sim \left(h_2-h_1 \right) \frac{\delta S}{\gamma c_\mathrm{v}}.
\end{equation}
So, the energy of the sound waves is directly proportional to the entropy change $\delta S$ and the difference in enthalpy $h_2-h_1$. If the flow is uniform ($h_2=h_1$), no sound waves are produced.

Assuming that the entropy fluctuations are due to dissipation of turbulent kinetic energy, the specific entropy can be estimated to an order of magnitude as
\begin{equation}
\delta S  \sim \frac{dQ}{T} \sim \frac{\delta \upsilon ^2}{T} 
\end{equation}
The temperature is related to the speed of sound $c$ as $T= \mu c^2 / (\gamma R)$, where $\mu$ is the molar mass and $R$ is the universal gas constant. Using this, we obtain
\begin{equation}
\delta S  \sim \gamma \frac{R}{\mu} \delta {\cal M}^2,
\end{equation}
where $\delta {\cal M}=\delta \upsilon / c$ is the turbulent Mach number. In the innermost nuclear burning shells, $\delta {\cal M} \sim 0.1$ \citep{collins:18}, which we adopt as our default value. since our formalism is linear, the results presented below can be rescaled to any chosen amplitude of entropy perturbations.

\begin{figure*}
    \centering
    \includegraphics[width=0.49\textwidth]{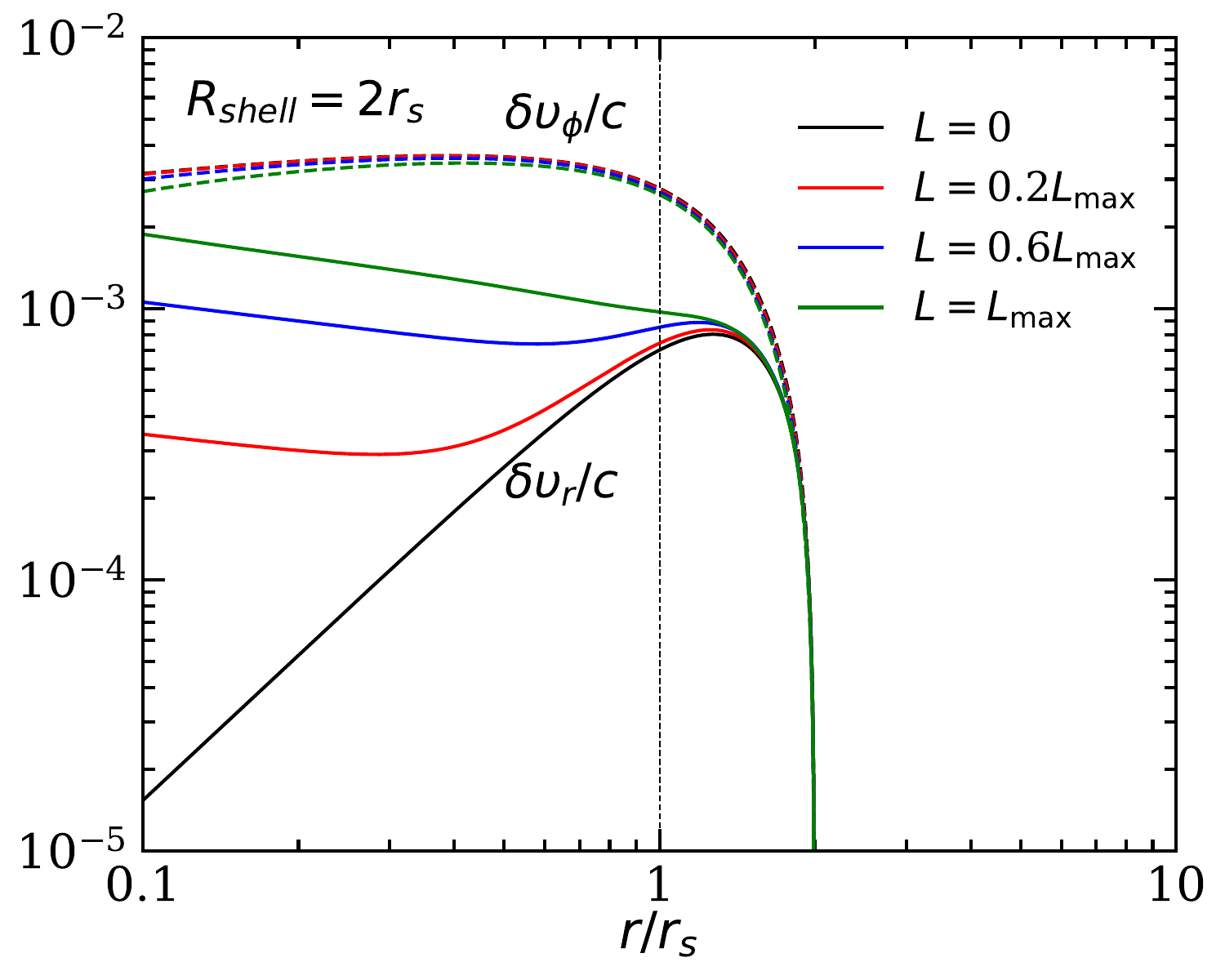}
    \includegraphics[width=0.49\textwidth]{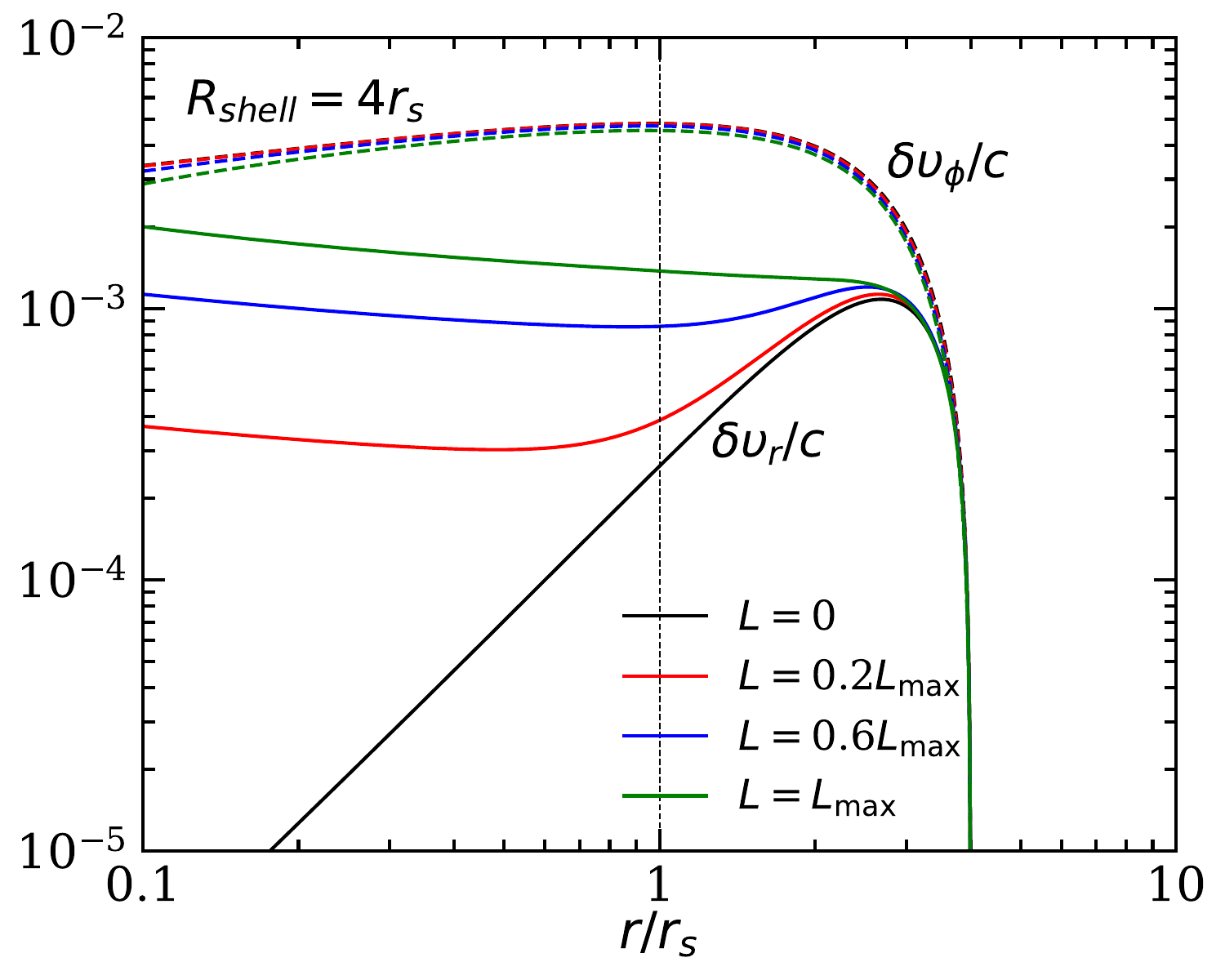}
    \caption{Radial and tangential velocities of vorticity produced by advected entropy perturbations versus radius for various values of the specific angular momenta $L$. Left and right panels reprensent the initial shell radii $R_{\rm shell}=2\rso$ and $R_{\rm shell}=4\rso$, respectively. The tangential component grows with rotation, whereas the radial component remains nearly unaffected.}
    \label{fig:vort_from_ent}
\end{figure*}

\begin{figure*}
    \centering
    \includegraphics[width=0.32\textwidth]{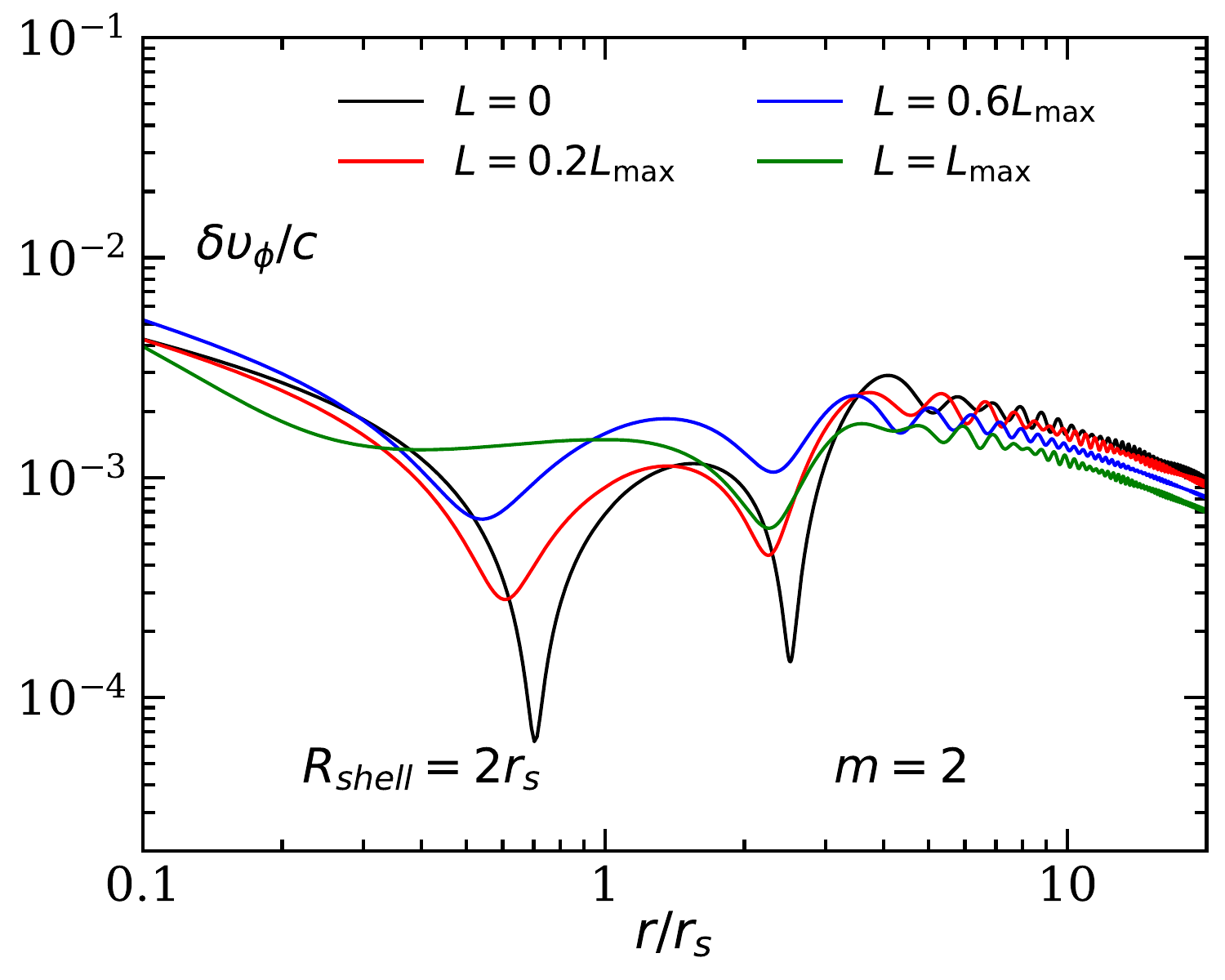}
    \includegraphics[width=0.32\textwidth]{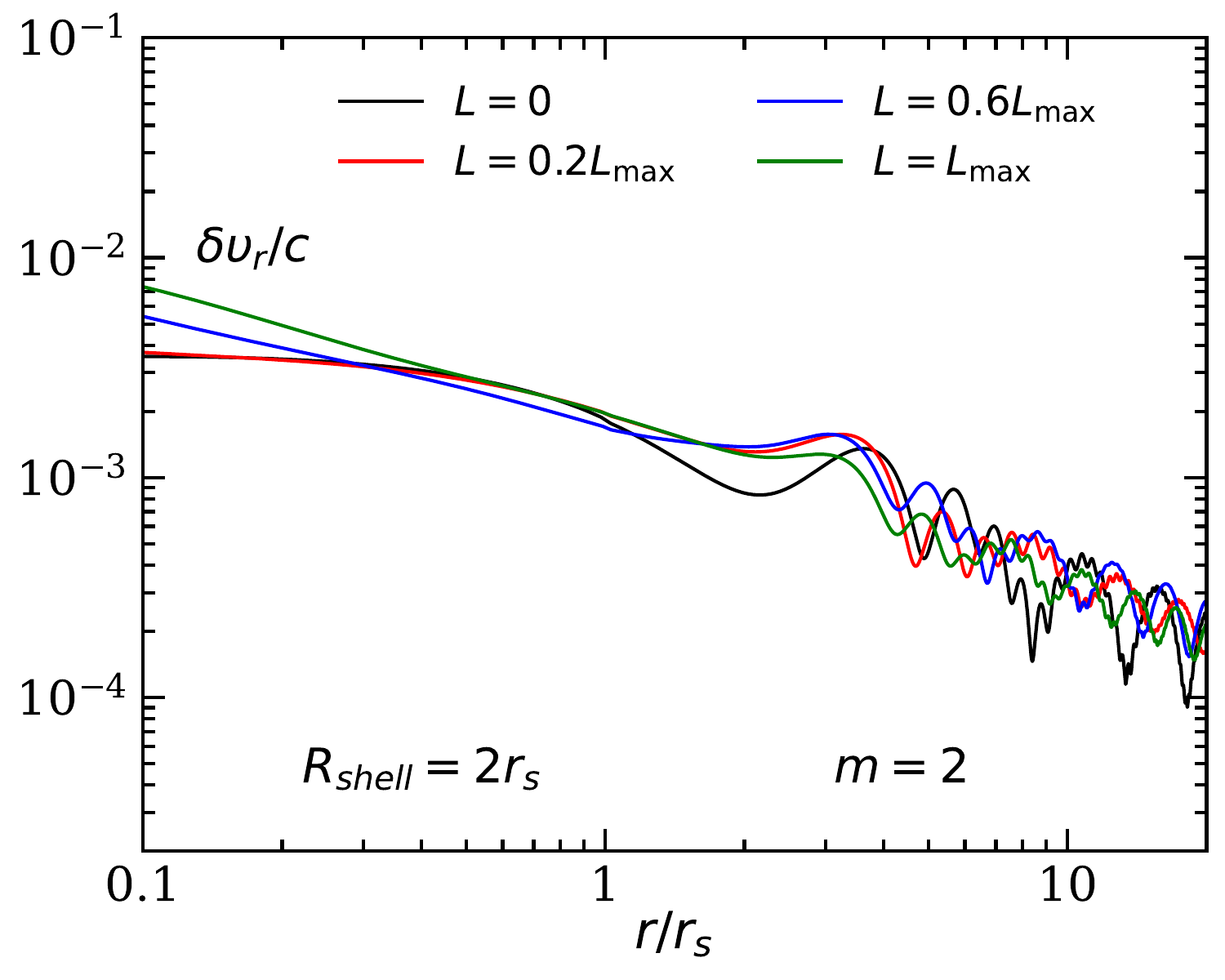}
    \includegraphics[width=0.32\textwidth]{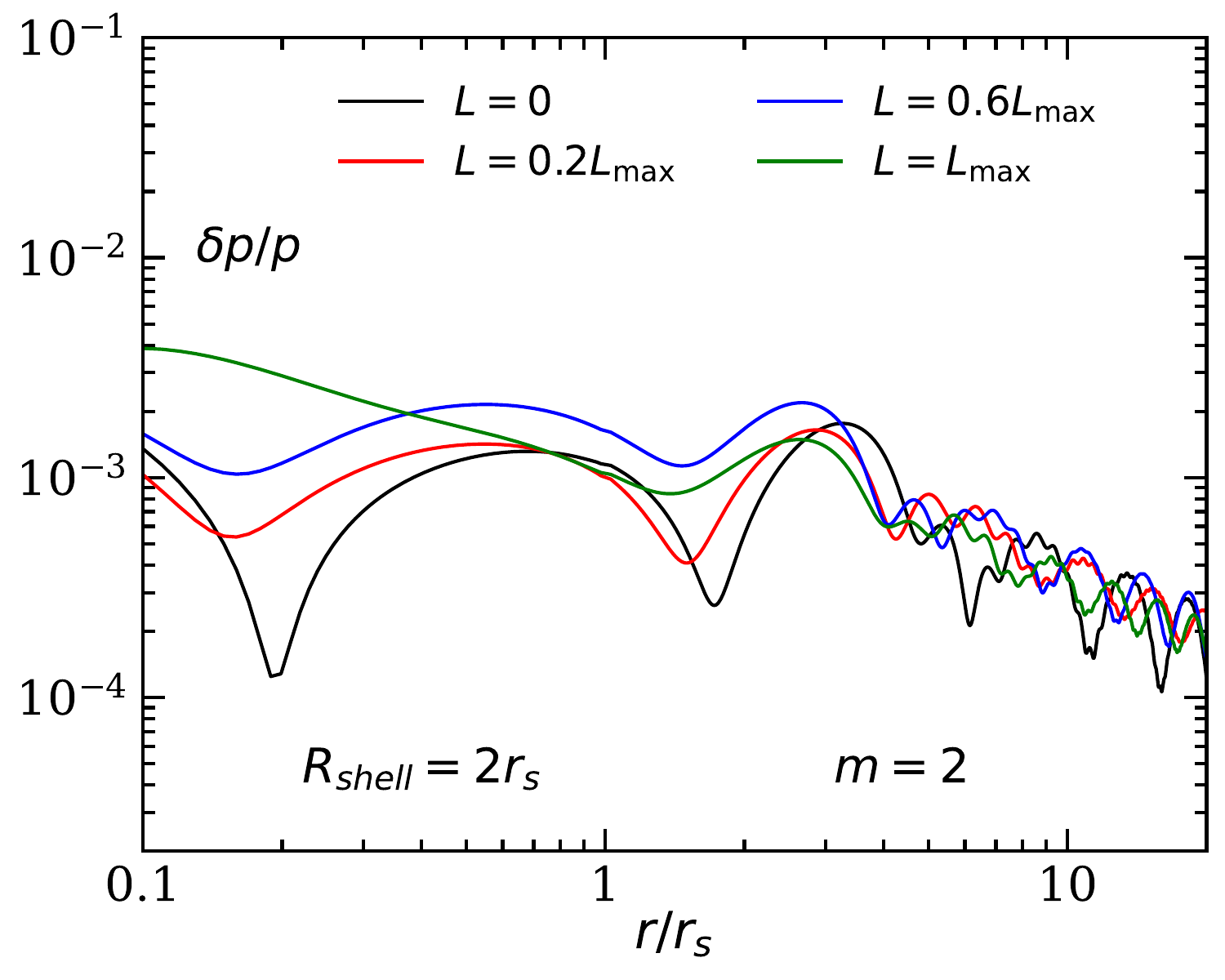}
    \caption{Velocity $\delta \upsilon_\phi/c$ (left panel) and $\delta \upsilon_r/c$ (center panel) as well as pressure perturbations (right panel) produced by entropy waves as a function of $r$ for a range of specific angular momentum values $L$ for the waves with $m=2$ and $R_\mathrm{shell}=2\rso$. }
    \label{fig:fullRepres}
\end{figure*}

\begin{figure*}
    \centering
    \includegraphics[width=0.32\textwidth]{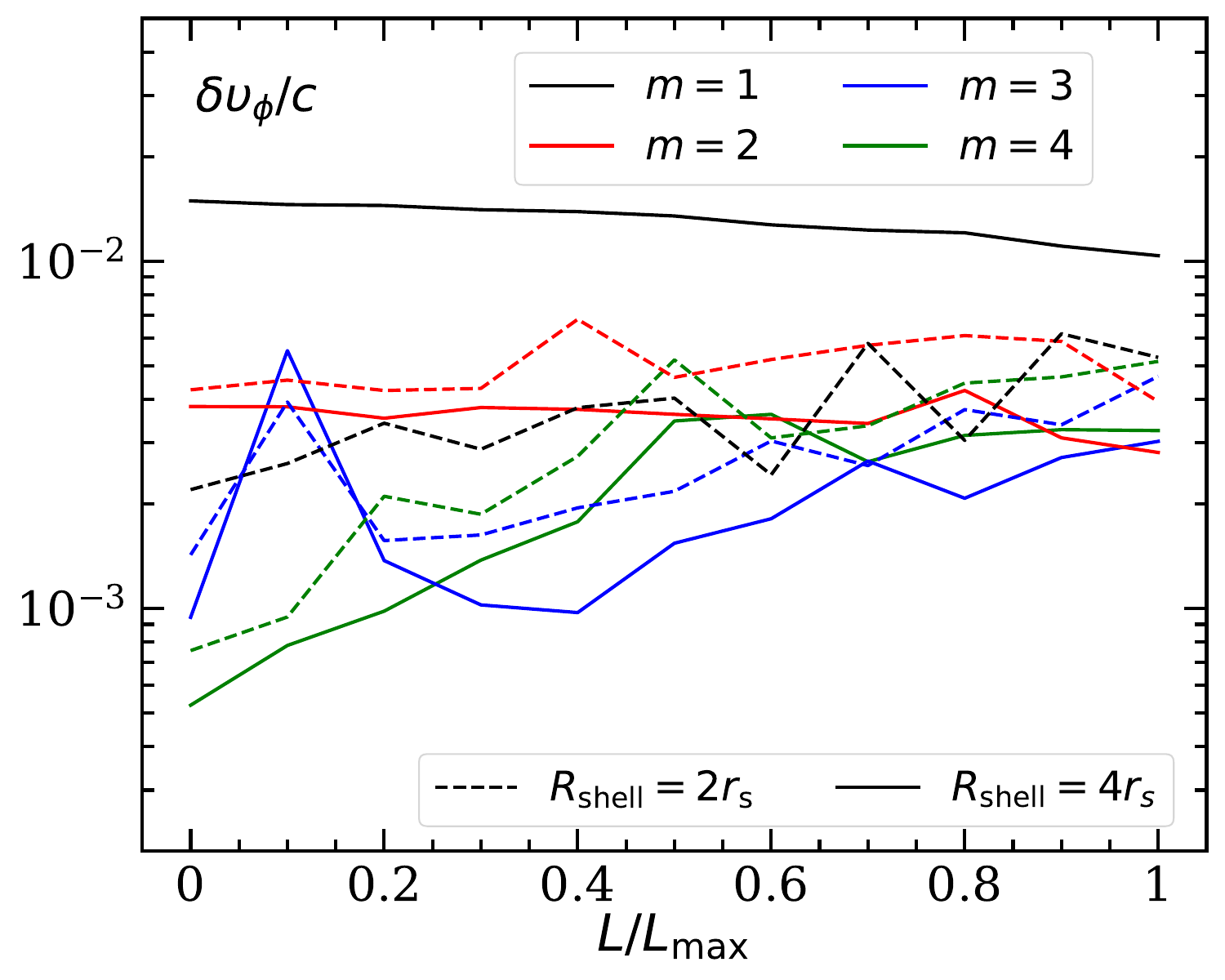}
    \includegraphics[width=0.32\textwidth]{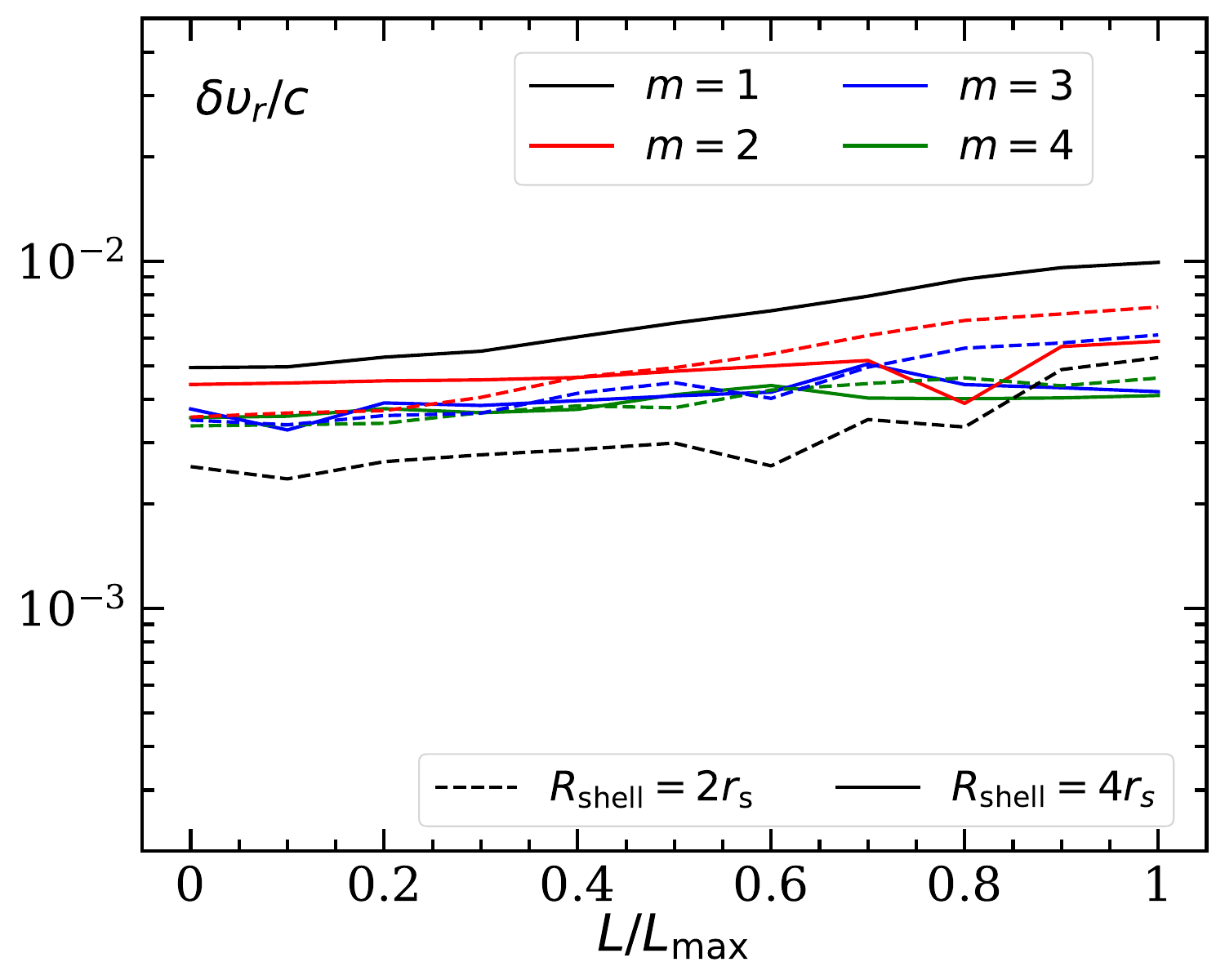}
    \includegraphics[width=0.32\textwidth]{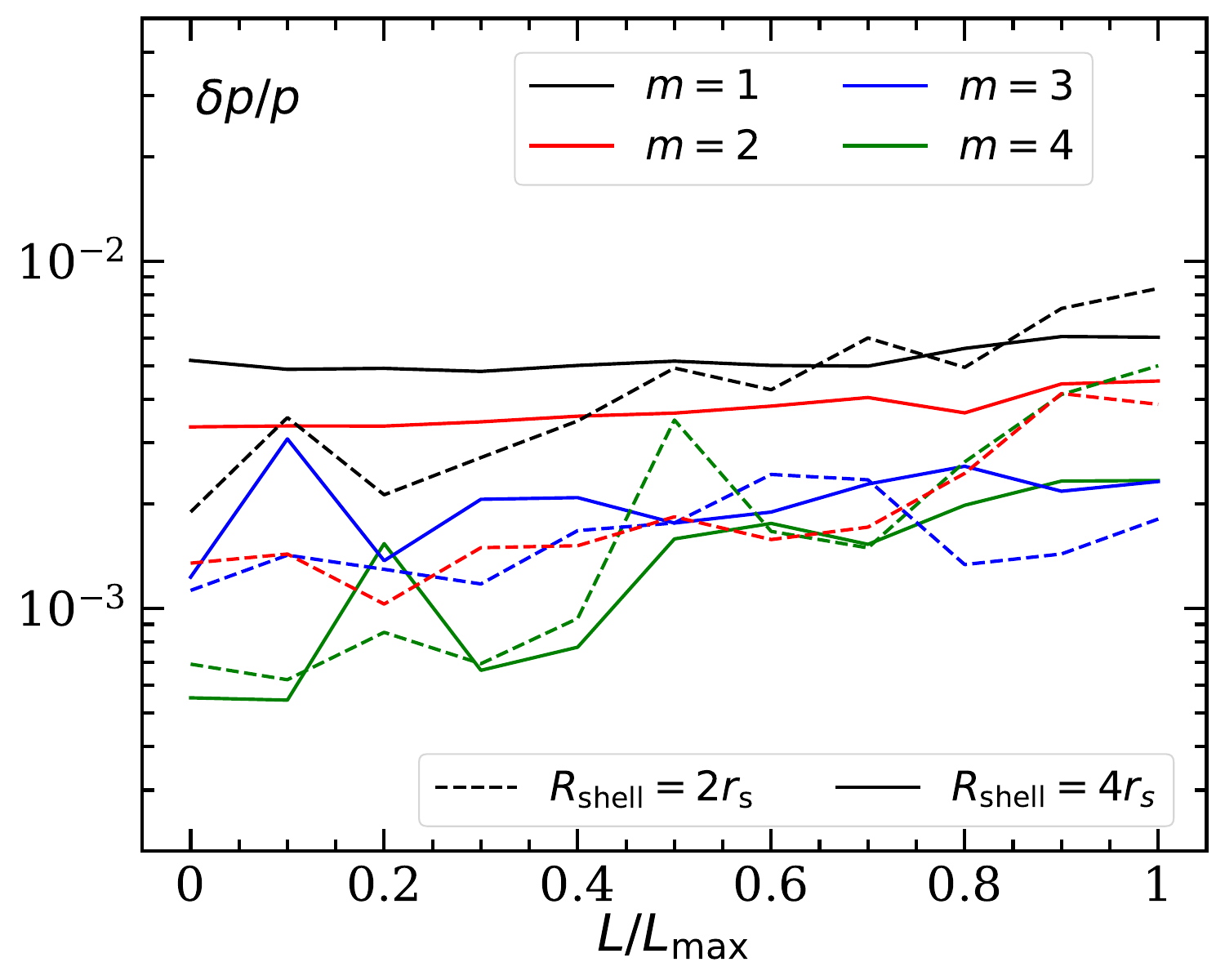}
    \caption{Velocity perturbations $\delta \upsilon_\phi/c$ (left panel) and $\delta \upsilon_r/c$ (center panel) and pressure perturbations $\delta p / p$ (right panel) produced by advected entropy waves at $R_{\rm shock} = 0.1r_s$ as function of the specific angular momentum $L$.}
    \label{fig:full_atRmin}
\end{figure*}

\begin{figure*}
    \centering
    \includegraphics[width=0.32\textwidth]{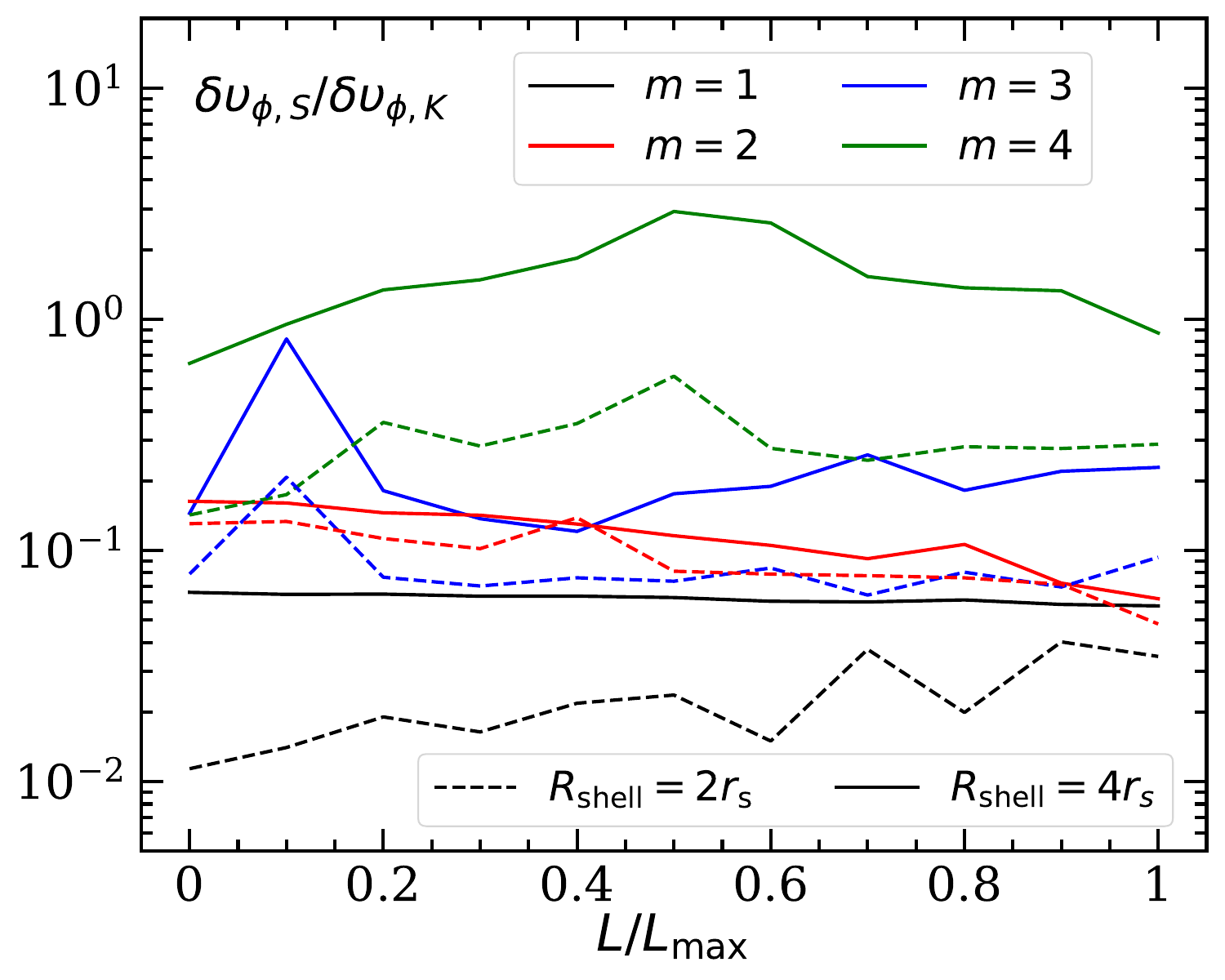}
    \includegraphics[width=0.32\textwidth]{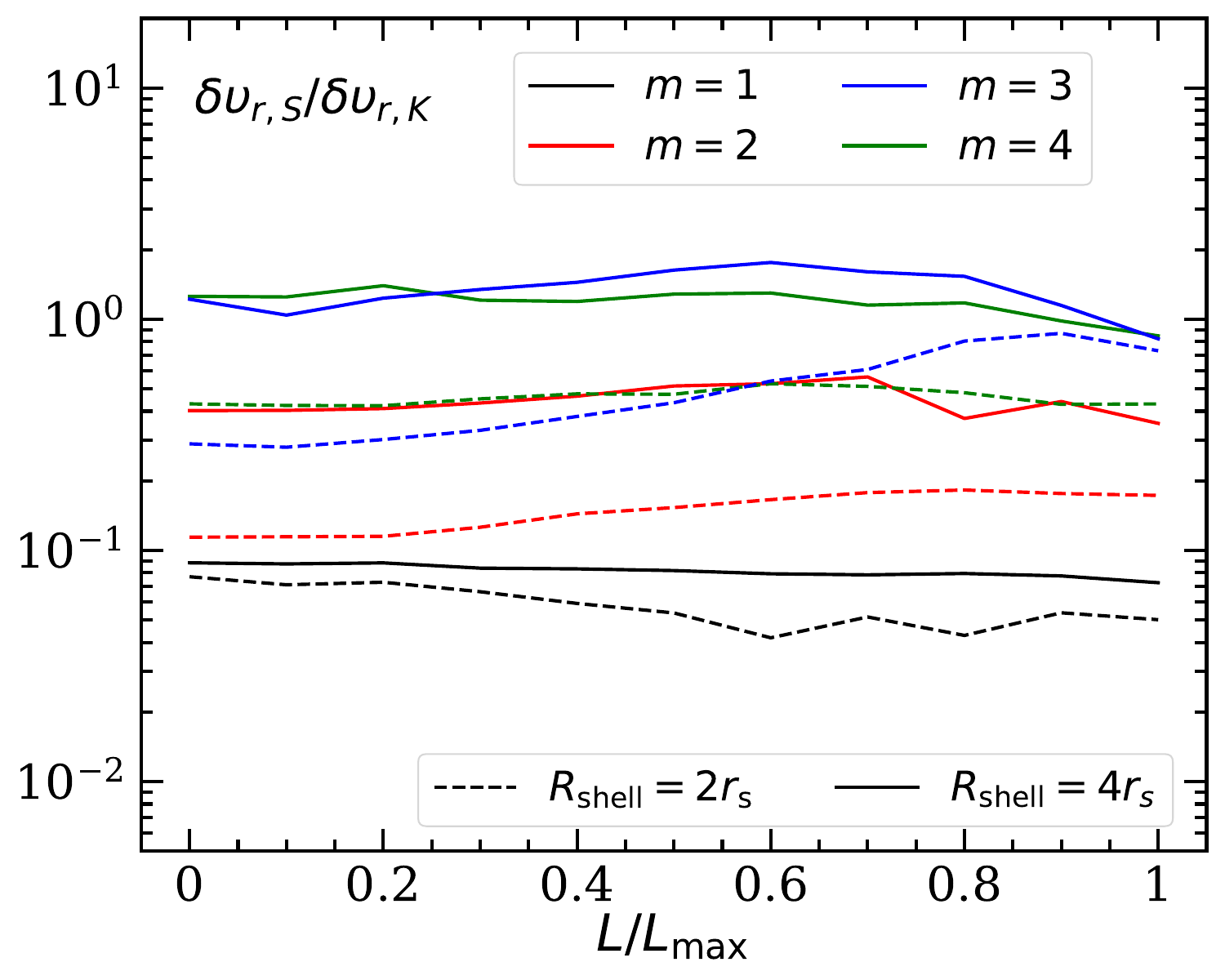}
    \includegraphics[width=0.32\textwidth]{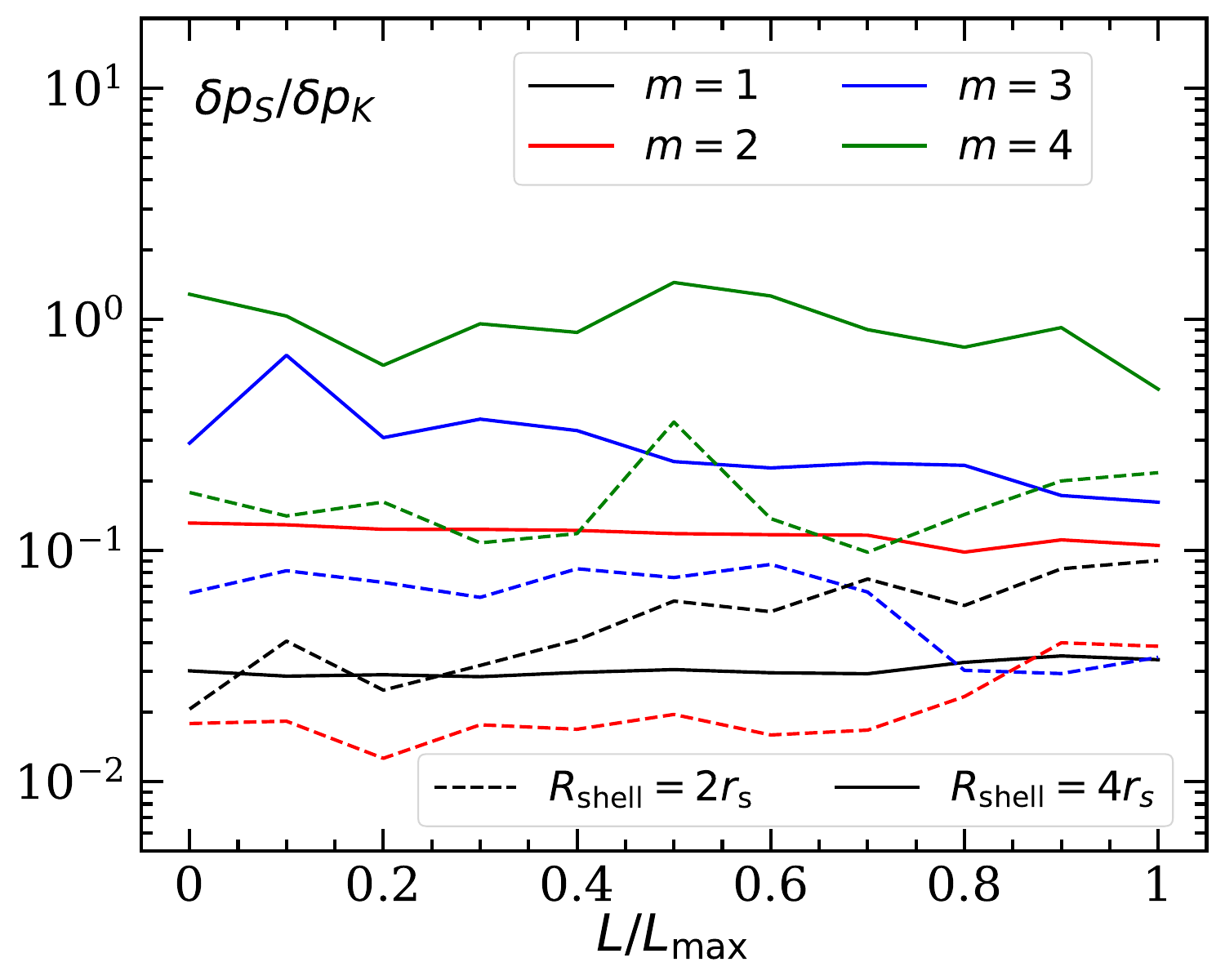}
    \caption{Ratio of velocity $\delta \upsilon_\phi/c$ (left panel) and $\delta \upsilon_r/c$ (center panel) as well as pressure $\delta p / p $ (right panel) perturbations produced by advected entropy waves and convective eddies at $R_{\rm shock} = 0.1r_s$ plotted against the specific angular momentum $L$. The convective eddies are scaled to a Mach number of $0.1$ prior to collapse.}
    \label{fig:full_atRmin_rat}
\end{figure*}

\subsection{Linear perturbative analysis}
\label{sec:quantitative}

Figure~\ref{fig:vort_from_ent} shows the radial (left panel) and azimuthal (right panel) components of the velocity of vorticity waves generated through the baroclinic effect of entropy perturbations, plotted as a function of radius for various values of the specific angular momentum $L$. The velocity perturbations are measured in units of local speed of sound (i.e., these components are similar to the convective Mach number components). We notice that the velocity perturbations never exceed $10^{-2}$ when they reach small radii ($\sim 0.1 r/\rso$). This radius corresponds to a few hundred kilometers in the early post-bounce phase, the location at which the perturbations are expected to meet the supernova shock \citep[e.g.,][]{mueller:17}. The radial component depends weakly on rotation, whereas the azimuthal ($\phi$) component increases with $L$ as the vorticity waves are advected inward for $m>1$ modes. This trend is straightforward to interpret: accelerated collapse stretches vortices radially, while faster rotation in the inner regions shears and distorts them in the azimuthal direction. Due to conservation of circulation, the stretching of voritices limits the velocities to low values, while the deformation in the azimuthal direction increases the $\phi$ velocity component. These effect is similar to the influence of rotation on convective eddies discussed in Paper I.

Figure~\ref{fig:fullRepres} shows the radial profiles of the velocity and pressure perturbations for $m=2$ and a range of specific angular momentum values $L$. The pressure perturbations are measured in units of the pressure of the background flow (i.e., $\delta p / p $). The velocity perturbations include contributions from both acoustic and vorticity waves generated by the advection of entropy perturbations \citep[e.g.,][]{foglizzo:01}. We notice that the acoustic wave amplitudes do not exceed $10^{-2}$ at any radius. Next, we investigate, how this depends on angular wavenumber $m$. 

Figure~\ref{fig:full_atRmin} shows the azimuthal and radial velocity as well as pressure perturbations for different values of $m$ and $R_\mathrm{shell}$ as a function of $L$ at a radius of $0.1\rso$, where these perturbations are anticipated to reach the supernova shock. The velocity perturbations include contributions from both acoustic and vorticity waves generated by the advection of entropy perturbations \citep[e.g.,][]{foglizzo:01}. The radial velocity perturbation, shown in the middle panel, is largely insensitive to rotation for all values of $m$ and $R_\mathrm{shell}$. Consequently, the pressure perturbations, arising from acoustic waves and shown in the right panel, are also not significantly affected by rotation. However, we observe an increase in the $\phi$ component of the velocity perturbation amplitude with rotation (left panel), particularly for $m=3$ and $m=4$. This behavior results from the deformation of perturbations in the radial direction due to the spin-up of stellar matter as it collapses toward smaller radii, as discussed above. 

Overall, the amplitudes of the velocity and pressure perturbations remain well below $0.01$, largely independent of rotation. For comparison, the convective eddies, which have typical Mach numbers of order $0.1$ \citep{collins:18}, are expected to generate perturbations of similar magnitude just ahead of the supernova shock for large scale modes \citep[e.g.,][]{mueller:16, abdikamalov20}. This indicates that the acoustic and vortical perturbations arising from the advection of entropy fluctuations are relatively weak. Consequently, their contribution to the dynamics of the explosion is expected to be sub-dominant compared to that of convective eddies themselves and other hydrodynamic instabilities \citep{mueller:15b, bruenn16, oconnor18, Matsumoto22, Burrows24}. 

To gain further insight into the relative contribution of entropy waves, Fig.~\ref{fig:full_atRmin_rat} shows the ratio of the contributions from entropy waves to those from the initial convective eddies (i.e., not generated by entropy waves) for the velocity perturbations $\delta \upsilon_\phi/c$ (left) and $\delta \upsilon_r/c$ (center), as well as pressure perturbations $\delta p / p $ (right), measured at $0.1 \rso$ plotted against the specific angular momentum $L$. The convective eddies are normalized to have the Mach number of $0.1$ at their origin. As expected, this ratio is largely insensitive to rotation. As we can see, the contribution of entropy waves grow relative to that of convective eddies with increasing $m$. This is expected as the acoustic waves generated by convective vortices scale as $\propto 1/m $ \citep{abdikamalov21}. For low-order modes ($m=1$ and $m=2$), the entropy contribution is smaller than that of the convective eddies. These large-scale modes are particularly important for explosion dynamics, as they strongly reduce the critical neutrino luminosity necessary to drive the supernova explosion \citep{Burrows93, mueller:16, collins:18}. Hence, we expect negligible contribution of entropy waves for these modes. For higher-order (smaller-scale) modes (e.g., $m=3$ and $m=4$), the entropy contribution can be comparable to that of convective eddies. However, since the lowering of the neutrino luminosity threshold for explosion scales as $\propto 1/m$ \citep[e.g.,][]{mueller:16, mueller:17}, these small-scale modes are expected to be less influential compared to the low-order, large-scale ones. This suggests that entropy waves play only a minor role in triggering the explosion\footnote{Note that the interaction of sound waves with the supernova shock produces entropy perturbations in the post-shock region \citep{abdikamalov18}, which can significantly influence the dynamics \citep{kazeroni20}. However, this is distinct from the effect we examine in this paper, namely the role of entropy waves originating in the pre-collapse star.}.

\section{Conclusion}
\label{sec:conclusion}

In this work, we analyzed how entropy disturbances evolve in massive, rotating stars during collapse. To do this, we used linear hydrodynamic equations on a steady-state background flow. The background flow was described with the rotating, transonic Bondi solution. The equations were then solved in spherical coordinates restricted to the equatorial plane (see Section~\ref{sec:method} for details). We study the effects of rotation over a wide range, from non-rotating models to stars rotating at extreme rates. 

We investigated vorticity waves generated by the advection of entropy perturbations through baroclinic effects. Our results show that the radial velocity component of these waves is largely unaffected by rotation, whereas the azimuthal component increases as the core spins faster. This behavior arises because collapse stretches vortices in the radial direction, while differential rotation shears and distorts them azimuthally. 

At small radii, where the perturbations are expected to impinge upon the supernova shock, we find that radial velocity and pressure perturbations remain largely insensitive to rotation. In contrast, the azimuthal velocity perturbations increase with rotation, especially for intermediate angular modes ($m=3,4$). This is a direct consequence of the spin-up of infalling matter as it moves inward. Despite this growth, the vortex velocities remain below one percent of the local sound speed, suggesting that their contribution to the explosion dynamics is minor (see Section~\ref{sec:results} for details). 

When comparing the contributions of entropy perturbations with those of the original convective eddies, we find that low-order modes ($m=1,2$) are dominated by convective eddies. These modes are most relevant for the explosion mechanism since they strongly reduce the critical neutrino luminosity required for shock revival. For higher-order modes ($m=3,4$), the contribution of entropy perturbations can become significant (relative to the contribution of the convective eddies), but these modes are expected to have a smaller impact on the explosion outcome \cite[e.g.,][]{mueller:16}. 
    
Our results provide insight into the evolution of entropy perturbations in rotating stellar cores prior to their encounter with the supernova shock. A key long-term objective is to determine how such pre-collapse perturbations influence the explosion dynamics in rotating models. Shock revival is governed by a number of intertwined processes, particularly the interaction of incoming acoustic waves with the shock front \citep[e.g.,][]{Huete20} and with the post-shock flow \citep[e.g.,][]{fernandez:15a, mueller:16}. While advanced three-dimensional simulations exist \citep[e.g.,][]{mueller:17, Vartanyan22}, the specific role of rotation in these interactions remains somewhat unexplored, even within simplified setups \citep[e.g.,][]{Buellet23}. 

It is important to emphasize that our analysis relies on an idealized framework constructed to isolate the influence of rotation from other competing effects. While this setup is not intended to yield precise quantitative predictions, it provides a powerful means of uncovering the fundamental mechanisms at play and enables a systematic parameter study. In this way, our work complements large-scale numerical simulations by offering a clearer physical interpretation of the processes that govern explosion dynamics.

\backmatter

\bmhead{Acknowledgements}

We thank Ernazar Abdikamalov, Thierry Foglizzo, and Sungbae Chun for useful discussions.

\bmhead{Data Availability}

The data used in this work can be shared on request to the corresponding author.

\begin{appendices}

\section{Mathematical Formalism}
\label{sec:formalism}

Our approach builds on the method developed in our previous work \citep{abdikamalov21}. For completeness, we summarize here the key elements, focusing on those most relevant to entropy perturbations.

Following Paper I, we adopt spherical coordinates $(r,\theta,\varphi)$ restricted to the equatorial plane $\theta=\pi/2$ and assume no variation in the poloidal direction. In the adiabatic approximation we adopt, the flow is described by the following hydrodynamics equations 
\begin{eqnarray}
{\p \rho\over \p t}+\boldsymbol{\nabla}\cdot\rho \boldsymbol{\upsilon} &=&0,\\
{\p  \boldsymbol{\upsilon}\over\p t}+\boldsymbol{w} \times \boldsymbol{\upsilon} +\boldsymbol{\nabla} \left({v^2\over 2}\!+\!{c^2\over\gamma-1}\!+\!\Phi_0\right)&=&{c^2\over\gamma}\boldsymbol{\nabla} S,\\
{\p S\over \p t}+\boldsymbol{\upsilon}\cdot\boldsymbol{\nabla} S&=&0.
\end{eqnarray}
The variables are defined as follows: $\upsilon$ is velocity, $\boldsymbol{w} = \nabla \times \boldsymbol{\upsilon}$ is vorticity of the flow, $\rho$ is density, $c$ is the speed of sound, $\gamma$ is the adiabatic index, and $S$ is entropy. The gravitational potential $\Phi_0$ is given by $\Phi_0\equiv -GM/r$, where $M$ is the neutron star mass. The stationary background solution is obtained as escribed in Appendix B of Paper I. Without loss of generality, we set $R/\mu = 1$ \citep[e.g.,][]{foglizzo:01}. 

After performing a Fourier transform in time, the resulting set of linearized equations is given by
\begin{flalign}
-i\omega\delta\rho+\boldsymbol{\nabla}\cdot(\rho \delta \boldsymbol{\upsilon}+\boldsymbol{\upsilon} \delta\rho)&=0,\\
-i\omega \delta \boldsymbol{\upsilon}+\delta \boldsymbol{w}\times \boldsymbol{\upsilon}+\boldsymbol{w}\times \delta \boldsymbol{\upsilon}
+\boldsymbol{\nabla}\left(\upsilon_r\delta \upsilon_r+\upsilon_\phi\delta \upsilon_\phi+{\delta c^2\over\gamma-1}\right)
&={c^2\over\gamma} \boldsymbol{\nabla} \delta S,\\
-i\omega \delta S+\boldsymbol{\upsilon}\cdot\boldsymbol{\nabla} \delta S&=0,
\end{flalign}
where $\omega$ is frequency. Following Paper I, we introduce variables $\delta f$, $\delta h$ as follows:
\begin{eqnarray}
\label{eq:df}
\delta f&\equiv&{\upsilon_r\delta \upsilon_r}+{L\over r}\delta \upsilon_\phi+ {\delta c^2\over\gamma-1},\\
\delta h&\equiv&{\delta \upsilon_r\over \upsilon_r}+{\delta \rho\over\rho}.
\end{eqnarray}
The linearized continuity equation in perturbed form yields
\begin{eqnarray}
-i\omega'\delta \rho +{1\over r^2}{\p\over \p r}{r^2\rho \upsilon_r\delta h}+{im\over r}{\rho \delta  \upsilon_\phi}=0,
\end{eqnarray}
where $\omega'$ is the Doppler shifted frequency,
\begin{eqnarray}
\omega'\equiv \omega-{mL\over r^2}.
\end{eqnarray}
The linearized Euler equation is expressed as:
\begin{eqnarray}
-i\omega\delta \upsilon_r +{L\over r}\delta w_\theta+{\p\delta f\over \p r}={c^2\over\gamma}{\p\delta S\over \p r},\\
-i\omega\delta \upsilon_\phi-\upsilon_r \delta w_\theta+{im\over r}\delta f={c^2\over\gamma}{im\over r}\delta S.\label{Euleraz2}
\end{eqnarray}
The entropy perturbation evolves according to
\begin{eqnarray}
\left({\p\over \p r}-{i\omega'\over \upsilon_r}\right)\delta S&=&0.\label{eqS2}
\end{eqnarray}
The curl of the Euler equation yields the vorticity equation :
\begin{eqnarray}
{\p\over \p r}(r\upsilon_r\delta w_\theta)={i\omega' r}\delta w_\theta-im{\delta S\over \gamma}{\p c^2\over \p r},
.\label{eqvort2}
\end{eqnarray}
where
\begin{eqnarray}
\delta w_\theta&=&-{1\over r}{\p r\delta \upsilon_\phi\over\p r}+{im\over r}\delta \upsilon_r.
\end{eqnarray}
is vorticity. The complete system can now be expressed through the following set of differential equations:
\begin{eqnarray}
{\p\delta f\over \p r}&=&i\omega\delta \upsilon_r +{c^2\over\gamma}{i\omega' \over \upsilon_r}\delta S-{L\over r}\delta w_\theta
,\\
{\p\delta h\over \p r}&=&{i\omega'\over \upsilon_r}{\delta \rho\over\rho} -{im\over r\upsilon_r}{\delta  \upsilon_\phi},\\
\left({\p\over \p r}-{i\omega'\over \upsilon_r}\right)\delta K&=&0,\\
\left({\p\over \p r}-{i\omega'\over \upsilon_r}\right)\delta S&=&0, \\ 
\label{eq:vtheta_f_dK1}
\omega r\delta \upsilon_\phi&=&-\frac{\delta K}{m} +{m}\delta f,\\
\label{eq:vtheta_f_dK2}
\delta f&=&{\omega\over m} r\delta \upsilon_\phi+{\delta K\over m^2},
\end{eqnarray}
where
\begin{eqnarray}
\delta K&\equiv &-im r\upsilon_r\delta w_\theta + m^2 {c^2\over\gamma}\delta S.
\end{eqnarray}
Quantities $\delta S$ and $\delta K$ can easily be integrated to
\begin{eqnarray}
\delta S&=&\delta S_0\e^{\int_{r_0} {i\omega'\over \upsilon_r}\d r},\\
\delta K&=&\delta K_0\e^{\int_{r_0} {i\omega'\over \upsilon_r}\d r},\\
\delta w_\theta&= & {i\over r\upsilon_r}\left\lbrack \frac{\delta K_0}{m} - m{c^2\over\gamma}\delta S_0\right\rbrack \e^{\int_{r_0} {i\omega'\over \upsilon_r}\d r}, \label{eq:wtheta}
\end{eqnarray}
If we express $\delta f$ in terms of $r\delta \upsilon_\phi$, we obtain
\begin{eqnarray}
{\p \over \p r}( r\delta \upsilon_\phi)&=&im\delta \upsilon_r - i {\delta K\over {m \upsilon_r}}+ im{c^2\over\gamma \upsilon_r}\delta S,\label{diff1}\\
{\p\delta h\over \p r}&=&{i\omega'\over \upsilon_r}{\delta \rho\over\rho} -{im\over r\upsilon_r}{\delta  \upsilon_\phi} \label{diff2}.
\end{eqnarray}
By introducing the new variable $X$, defined via,
\begin{eqnarray}
\label{eq:X}
{\p X\over \p r}\equiv {\upsilon_r\over 1-\M^2},
\end{eqnarray}
one can obtain 
\begin{flalign}
\label{eq:du_system}
\left({\p \over \p X}+{i\omega'\over c^2}\right)( r\delta \upsilon_\phi) &= im \delta h
 -i{\delta K\over {m\upsilon_r^2}}+{im\over \gamma}\delta S\left({1\over \M^2}+\gamma-1\right), \\
\label{eq:dh_system}
\left({\p \over \p X}+{i\omega'\over c^2}\right)\delta h &= {i W\over m} r\delta \upsilon_\phi -{i\omega'\over \upsilon_r^2} \left(\delta S-{\delta K\over m^2c^2}\right),
\end{flalign}
The parameter $W$ is
\begin{eqnarray}
W = \frac{\omega'^2\mu^2}{\upsilon_r^2 c^2},
\end{eqnarray}
where
\begin{eqnarray}
\mu^2\equiv 1-{m^2c^2\over r^2\omega'^2}(1-\M^2).
\end{eqnarray}
Combining these together, we obtain a single 2nd-order ordinary differential equation
\begin{flalign}
\left\lbrace\left({\p \over \p X}+{i\omega'\over c^2}\right)^2+W \right\rbrace & (r\delta \upsilon_\phi) = {\omega'm\over \upsilon_r^2}
\left(\delta S-{\delta K\over m^2c^2}\right) \nonumber \\
&-im\left({\p \over \p X}+{i\omega'\over c^2}\right)\left\lbrack
{\delta K\over m^2\upsilon_r^2}
-{\delta S\over \gamma}\left({1\over \M^2}+\gamma-1\right)\right\rbrack.
\end{flalign}
Using Eq.~(\ref{eq:wtheta}), the equation can be expressed as
\begin{eqnarray}
\left\lbrace\left({\p \over \p X}+{i\omega'\over c^2}\right)^2+W \right\rbrace( r\delta \upsilon_\phi)=
-{\p \over \p X}\left({r\delta w_\theta\over \upsilon_r}
\right),\label{eq:diff_eq1}
\end{eqnarray}
This equation closely resembles that of the corresponding flow without rotation. The role of angular momentum enters only through $\omega'$ \citep{yamasaki:08}. We now introduce the variable
\begin{flalign}
\delta \tilde \upsilon_\phi\equiv \delta  \upsilon_\phi \e^{\int {i\omega'\over c^2}\d X}, \label{eq:vphitilde}
\end{flalign}
in terms of which, the system of Eqs.~(\ref{eq:du_system})–(\ref{eq:dh_system}) can be rewritten as:
\begin{flalign}
\label{eq:dut_system}
{\p (r\delta \tilde \upsilon_\phi) \over \p X} &= im \delta \tilde h + \left[-i{\delta K\over {m\upsilon_r^2}}+{im\over \gamma}\delta S\left({1\over \M^2}+\gamma-1\right) \right] \e^{\int {i\omega'\over c^2}\d X}, \\
\label{eq:dht_system}
{\p \delta \tilde h \over \p X} &= {i W\over m} r\delta \tilde \upsilon_\phi -{i\omega'\over \upsilon_r^2} \left(\delta S-{\delta K\over m^2c^2}\right) \e^{\int {i\omega'\over c^2}\d X}.
\end{flalign}
Eq.~(\ref{eq:diff_eq1}) can be rewritten as:
\begin{flalign}
\left\lbrace{\p^2 \over \p X^2}+W\right\rbrace( r\delta \tilde \upsilon_\phi)=-
\e^{\int {i\omega'\over c^2}\d X}{\p \over \p X}{r\delta w_\theta\over \upsilon_r}. \label{rdvtilde} 
\end{flalign}
The in-going and out-going waves, which represent the solution of the homogeneous part of the linearized equations, as well as boundary conditions are obtained following the method outlined in Section C1 of Paper I. Once we know $\delta f$, $\delta h$, $\delta \upsilon_\phi$, $\delta K$, and $\delta S$, we can obtain other hydrodynamic variables via expressions 
\begin{eqnarray}
{\delta \upsilon_r\over \upsilon_r}&=&{1\over 1-\M^2}\left(\delta h+\delta S-{\omega'\over mc^2} r\delta \upsilon_\phi-{\delta K\over m^2c^2} \right), \label{dv/v}\\
{\delta \rho\over\rho}&=&-{1\over 1-\M^2}\left(\M^2\delta h+\delta S-{\omega'\over mc^2} r\delta \upsilon_\phi-{\delta K\over m^2c^2}\right), \label{drho/rho}\\
{\delta p\over {\gamma p}}&=&-{1\over 1-\M^2} \left( \M^2(\delta h+\delta S)+(1-\M^2){\delta S\over\gamma}-{\omega'\over mc^2}r\delta \upsilon_\phi-{\delta K\over m^2c^2}\right). \label{dp/p}
\end{eqnarray}

\subsection{Inhomogeneous solution}
\label{sec:full_inho_sol}

To obtain the full solution generated by advected entropy perturbations, we linearly superpose two solutions: one with $dS$ and another with $\delta K = m^2 c_\mathrm{shell}^2{\delta S}/{\gamma}$, where $c_\mathrm{shell}$ is the speed of sound at the initial radius of the perturbations. The latter solution takes into account the contribution of vortices generated by advected entropy perturbations. This choice $\delta K = m^2 c_\mathrm{shell}^2{\delta S}/{\gamma}$ guarantees vanishing vorticity at the initial radius of the perturbations. 

To derive the solution for advected vorticity waves ($\delta S \ne 0$, $\delta K=0$), we begin by rewriting Eq.~(\ref{rdvtilde}) in terms of $\delta \tilde f$:
\begin{flalign}
\left\lbrace{\p^2 \over \p X^2}+W\right\rbrace \delta\tilde f \!=\!-\frac{1\!-\!{\cal M}^2}{\upsilon_r}
\e^{\int {i\omega'\over c^2}\d X}  \frac{\omega}{{\cal M}^2} \left(i\partial_r \! \log {\cal M}^2 \!+\! \frac{\omega'}{\upsilon_r}\right) \frac{\delta S}{\gamma}. \label{dftilde} 
\end{flalign}
We solve this equation by generalizing the method used for the non-rotating case \citep{foglizzo:01} to non-uniform frequency $\omega'$, as proposed by \citet{yamasaki:08}. Since 
\begin{equation}
    \e^{\int {i\omega'\over c^2}\d X} \delta S = \e^{\int {i\omega'\over c^2}\d X} \e^{\int {i\omega'\over \upsilon_r} \d r} \delta S_R = \e^{\int {i\omega'\over \upsilon_r^2} \d X} \delta S_R,
\end{equation}
we can write Eq.~(\ref{dftilde}) as 
\begin{flalign}
\left\lbrace{\p^2 \over \p X^2}+W\right\rbrace \delta\tilde f \!=\!-\frac{1\!-\!{\cal M}^2}{\upsilon_r}
\e^{\int {i\omega'\over \upsilon_r^2}\d X}  \frac{\omega}{{\cal M}^2} \left(i\partial_r \! \log {\cal M}^2 \!+\! \frac{\omega'}{\upsilon_r}\right) \frac{\delta S_R}{\gamma}. \label{dftilde2} 
\end{flalign}
Assuming regularity at the sonic point and no incoming acoustic waves at infinity, we obtain
\begin{flalign}
\delta f(r>\rso)=-{\delta S_R\over2\gamma} &\bigg\{\delta f^-
\int_{\rso}^r
{\delta \tilde f_0}{\p\over\p r}
\left( {1-\M^2\over \M^2}\e^{\int_R^r{{i\omega' \d r}\over \upsilon_r(1-\M^2)}}
\right)\d r \nonumber \\
&-\delta f_0
\int_{\infty}^r
{\delta \tilde f^-}
{\p\over\p r}
\left( {1-\M^2\over \M^2}
\e^{i\omega\int_R^r{{i\omega' \d r} \over \upsilon_r(1-\M^2)}}
\right)
\d r
\bigg\},
\end{flalign}
where $\delta f^-$ and $f_0$ are homogeneous solutions and they are obtained as described in Appendix C1 of Paper I. After an integration by parts, we obtain the following
\begin{flalign}
\delta f(r>\rso)={\delta S_R\over2\gamma} \bigg\{&\delta f^-
\int_{\rso}^r
{\p{\delta \tilde f_0}\over\p r}
{1-\M^2\over \M^2}\e^{\int_R^r{{i\omega' \d r} \over \upsilon_r(1-\M^2)}}
\d r \nonumber \\
&-\delta f_0
\int_{\infty}^r
{\p{\delta \tilde f^-}\over\p r}
{1-\M^2\over \M^2}\e^{\int_R^r{{i\omega' \d r} \over \upsilon_r(1-\M^2)}}
\d r
\bigg\}.\label{eq:fsub1}
\end{flalign}
To simplify this further, we re-write the system (\ref{eq:dut_system})-(\ref{eq:dht_system}) into a system for homogeneous functions $\delta \tilde f$ and $\delta h$:
\begin{eqnarray}
\label{eq:dft_system_hom}
{\p \delta \tilde f \over \p X} &=& i \omega \delta \tilde h,  \\
\label{eq:dht_system_hom}
{\p \delta \tilde h \over \p X} &=& {i W\over \omega} \delta \tilde f.
\end{eqnarray}
where we used relation $r\delta\upsilon_\phi=(m/\omega) \delta f$ for homogeneous solution (cf. Eq.~\ref{eq:vtheta_f_dK1}). Since
\begin{equation}
    \frac{\d}{\d X} = \frac{1-{\cal M}^2}{\upsilon_r} \frac{\d}{\d r}, 
\end{equation}
the above system can be written as
\begin{eqnarray}
\label{eq:dft_system_hom}
{\p \delta \tilde f \over \p r} &=& \frac{i \omega \upsilon_r}{1-{\cal M}^2} \delta \tilde h,  \\
\label{eq:dht_system_hom}
{\p \delta \tilde h \over \p r} &=& {i W\over \omega} \frac{\upsilon_r}{1-{\cal M}^2} \delta \tilde f.
\end{eqnarray}
Using this, we can re-write solution (\ref{eq:fsub1}) as 
\begin{flalign}
\delta f(r>\rso)={{i \omega \delta S_R}\over2\gamma} \bigg\{&\delta f^-
\int_{\rso}^r
{{\delta \tilde h_0}}
{\upsilon_r \over \M^2}\e^{\int_R^r{{i\omega' \d r} \over \upsilon_r(1-\M^2)}}
\d r \nonumber \\
&-\delta f_0
\int_{\infty}^r
{\delta \tilde h^-}
{\upsilon_r\over \M^2}\e^{\int_R^r{{i\omega' \d r} \over \upsilon_r(1-\M^2)}}
\d r
\bigg\}.\label{eq:fsub2}
\end{flalign}
One more integration by parts yields integrals that converge at infinity:
\begin{flalign}
\delta f(r>\rso)&={{\delta S_R}\over2\gamma} 
 \bigg\{ \frac{2\omega}{\omega'}(c^2-\upsilon_r^2) e^{\int\frac{i\omega'}{\upsilon_r}\d r} \nonumber \\
&-\delta f^-
\int_{\rso}^r
\p_r \left[\delta \tilde h_0 (c^2-\upsilon_r^2)\frac{\omega}{\omega'} \right ]
\e^{\int_R^r{{i\omega' \d r} \over \upsilon_r(1-\M^2)}}
\d r \nonumber \\
&+\delta f_0
\int_{\infty}^r
\p_r \left[\delta {\tilde h}^- (c^2-\upsilon_r^2)\frac{\omega}{\omega'} \right ]
\e^{\int_R^r{{i\omega' \d r} \over \upsilon_r(1-\M^2)}}
\d r
\bigg\}. \label{eq:fsub3}
\end{flalign}
We write it in a more compact form:
\begin{flalign}
\delta f(r>\rso)&={{\delta S_R}\over2\gamma} 
 \bigg\{ \frac{2\omega}{\omega'}(c^2-\upsilon_r^2) e^{\int\frac{i\omega'}{\upsilon_r}\d r} \nonumber \\
&-\delta f^-
\int_{\rso}^r
G(r)
\e^{\int_R^r{{i\omega' (1+\M^2) \d r} \over \upsilon_r(1-\M^2)}}
\d r \nonumber \\
&+\delta f_0
\int_{\infty}^r
H(r)
\e^{\int_R^r{{i\omega' (1+\M^2) \d r} \over \upsilon_r(1-\M^2)}}
\d r
\bigg\}, \label{eq:fsub4}
\end{flalign}
where
\begin{eqnarray}
    G(r) = \frac{i\upsilon_r c^2 W}{\omega'} \delta f_0 + \frac{\omega}{\omega'} \bigg[\frac{\p c^2}{\p r} - \frac{\p \upsilon_r^2}{\p r} - (c^2-\upsilon_r^2) \frac{2 mL}{\omega' r^3} \bigg] \delta h_0
\end{eqnarray}
and
\begin{eqnarray}
    H(r) = \frac{i\upsilon_r c^2 W}{\omega'} \delta f^- + \frac{\omega}{\omega'} \bigg[\frac{\p c^2}{\p r} - \frac{\p \upsilon_r^2}{\p r} - (c^2-\upsilon_r^2) \frac{2 mL}{\omega' r^3} \bigg] \delta h^-.
\end{eqnarray}
One drawback of integration by parts is that it introduces terms $\propto 1/\omega'$, which become singular at the corotation point, where the Doppler-shifted frequency $\omega'$ becomes zero. We avoid this singularity by combining integration with and without integration by parts around that point. 

The solution for function $\delta h$ is obtained using Eq.~(\ref{eq:dut_system}):
\begin{flalign}
\delta h(r>\rso)={{\delta S_R}\over2\gamma} 
 \bigg\{ &- \delta h^-
\int_{\rso}^r
G(r)
\e^{\int_R^r{{i\omega' (1+\M^2) \d r} \over \upsilon_r(1-\M^2)}}
\d r \nonumber \\
&-\delta h_0
\int_{\infty}^r
H(r)
\e^{\int_R^r{{i\omega' (1+\M^2) \d r} \over \upsilon_r(1-\M^2)}}
\d r - 2 \gamma  \e^{\int\frac{i\omega'}{\upsilon_r}\d r} 
\bigg\}, \label{eq:hsub4}
\end{flalign}
where $\delta h^-$ and $\delta h_0$ are homogeneous solutions. In the supersonic domain ($r<\rso$), the homogeneous solution is constructed using $\delta f$ together with the Wronskian \citep{abdikamalov20}. The inhomogeneous solution is subsequently derived using an approach analogous to that employed for the subsonic region.

\subsection{Generation of vorticity by advected entropy waves}
\label{sec:entropy}

To calculate the vorticity generated by advected entropy perturbations, we linearly superpose two solutions: one with $dS$ and another with $\delta K = m^2 c_\mathrm{shell}^2{\delta S}/{\gamma}$, where $c_\mathrm{shell}$ is the speed of sound at the initial radius of the perturbations. This choice guarantees that the vorticity at the initial radius is zero. 

To obtain vortex velocities, we use the incompressibility condition described in Appendix F of \citet{abdikamalov20}. We rewrite Eq.~(\ref{rdvtilde}) as
\begin{flalign}
\left(\frac{\p^2}{\p X^2} + W\right)(r\delta \tilde{\upsilon}_\phi) = A\delta S_0\e^{\int \frac{i\omega' dX}{\upsilon_r^2}},
\end{flalign}
where $A \equiv m \omega'(1-\M^2) \left(c_\mathrm{shell}^2-{c^2}\right)/{\gamma}{\upsilon_r^4}$. The solution is
\begin{flalign}
r\delta \tilde{\upsilon}_\phi = {A \over W-\frac{\omega'^2}{\upsilon_r^4}}\delta S_0 \e^{\int\frac{i\omega'dX}{\upsilon_r^2}},
\end{flalign}
or
\begin{flalign}
r\delta \upsilon_\phi = r\delta\tilde{\upsilon}_\phi \e^{-\int\frac{i\omega'dX}{c^2}} = {A\over W-\frac{\omega'^2}{\upsilon_r^4}}\delta S,
\end{flalign}
from which we can obtain
\begin{flalign}
\frac{\delta \upsilon_\phi}{\upsilon_r} &= \frac{m}{\omega' \upsilon_r r}\frac{1-\M^2}{1-\mu^2\M^2}\left(c^2-c^2_\mathrm{shell}\right)\frac{\delta S}{\gamma}.
\end{flalign}
From Eq.~(\ref{diff1}), we obtain
\begin{equation}
    {\delta \upsilon_r\over \upsilon_r} = \frac{\omega'}{m \upsilon_r^2}r\delta{\upsilon_\phi} + \frac{\delta K}{m^2 \upsilon_r^2} - \frac{\delta S}{\gamma \M^2},
\end{equation}
where we replaced the radial derivative with a multiplication by a factor $i\omega'/\upsilon_r$, which is valid for advected perturbations. After substitution of the value of $\delta \upsilon_\phi$ and $\delta K = m^2 c_\mathrm{shell}^2{\delta S}/{\gamma}$, we obtain the following equation:
\begin{flalign}
    \frac{\delta \upsilon_r}{\upsilon_r} &= \frac{1-\mu^2}{1-\mu^2\M^2}\left(\frac{c^2_\mathrm{shell}}{c^2}-1\right)\frac{\delta S}{\gamma}.
\end{flalign}
Using Eqs. (\ref{eq:vtheta_f_dK2}) and (\ref{eq:dh_system}) leads to
\begin{flalign}
    \delta f^S &= \left[ \frac{\omega}{\omega'} \frac{1-\M^2}{1-\mu^2\M^2}(c^2-c_\mathrm{shell}^2) + c_\mathrm{shell}^2 \right] \frac{\delta S}{\gamma}, \\
    \delta h^S &= \left[ \frac{\mu^2(1-\M^2)}{1-\mu^2\M^2}\left(1-\frac{c_\mathrm{shell}^2}{c^2}\right) + \frac{c_\mathrm{shell}^2}{c^2} - \gamma \right] \frac{\delta S}{\gamma}.
\end{flalign}

\end{appendices}

\bibliography{sn-bibliography}

\end{document}